\documentclass[12pt]{article}

\usepackage{amsmath}
\usepackage{amssymb}
\usepackage{cite}

\usepackage{graphics}
\usepackage{graphicx}
\usepackage{color}
\usepackage{here}
\usepackage[hyphens]{url}
\usepackage{hyperref}
\usepackage[hyphenbreaks]{breakurl}
\usepackage{authblk}

\makeatletter
\renewcommand{\@biblabel}[1]{\quad#1.}
\makeatother

\title{Win-stay lose-shift strategy in formation changes in football}
\author[1,2]{Kohei Tamura}
\author[3,*]{Naoki Masuda}
\affil[1]{Department of Creative Informatics, The University of Tokyo, Tokyo, Japan}
\affil[2]{CREST, JST, Saitama, Japan}
\affil[3]{Department of Engineering Mathematics, University of Bristol, Bristol, UK}
\affil[*]{Corresponding author (naoki.masuda@bristol.ac.uk)}
\date{}

\begin{document}
\maketitle

\newpage
\section*{Abstract}
Managerial decision making is likely to be a dominant determinant of performance of teams in team sports. 
Here we use Japanese and German football data to investigate correlates between temporal patterns of formation changes across matches and match results. 
We found that individual teams and managers both showed win-stay lose-shift behavior, a type of reinforcement learning. 
In other words, they tended to stick to the current formation after a win and switch to a different formation after a loss. 
In addition, formation changes did not statistically improve the results of succeeding matches.
The results indicate that a swift implementation of a new formation in the win-stay lose-shift manner may not be a successful managerial rule of thumb.

\section{Introduction}
Exploring rules governing decision making has been fascinating various fields of research, and its domain of implication ranges from our daily lives to corporate and governmental scenes.
In economic contexts in a widest sense, individuals often modify their behavior based on their past experiences, attempting to enhance the benefit received in the future.
Such decision making strategies are generally called reinforcement learning. 
In reinforcement learning, behavior that has led to a large reward will be selected with a larger frequency, or the behavior will be incrementally modified toward the rewarded one. 
Reinforcement learning is common in humans \cite{Fudenberg1998, Camerer2003} and non-humans \cite{Pearce2013}, is implemented with various algorithms \cite{Sutton1998}, has theoretical underpinnings \cite{Fudenberg1998, Sutton1998}, and has neural substrates \cite{Schultz1997, Glimcher2009}. 

A simple version of reinforcement learning is the so-called win-stay lose-shift (WSLS) strategy \cite{Kraines1989, Nowak1993}.
An agent adopting this strategy sticks to the current behavior if the agent is satisfied.
The agent changes its behavior if unsatisfied.  
Experimental studies employing human participants have provided a line of evidence in favor of WSLS in situations such as repeated Prisoner's Dilemma \cite{Wedekind1996, Milinski1998}, gambling tasks \cite{Hayden2009, Scheibehenne2011}, and tasks in which participants construct virtual stone tools \cite{Mesoudi2008a, Mesoudi2008b, Mesoudi2014}. 
It has also been suggested in nonscientific contexts that decisions by athletes and gamblers are often consistent with WSLS patterns even if the outcome of games seems to be independent of the decision \cite{Vyse2013}. 

Association football (also known as soccer; hereafter refer to it as football) is one of the most popular sports in the world and provides huge business opportunities.
The television rights of the English Premier League yield over two billion euros per year \cite{Panja2013}.
Transfer fees of top players can be tens of millions of euros \cite{Dobson2011}.
Various aspects of football, not only watching but also betting \cite{Dixon1997} and the history of tactics \cite{Wilson2013}, enjoy popularity.
Football and other team sports also provide data for leadership studies because a large amount of sports data is available and the performance of teams and players can be unambiguously measured by match results \cite{Audas1997, Dawson2000, Audas2002}.

In the present study, using data obtained from football matches, we examine the possibility that managers of teams use the WSLS strategy.
Managers can affect the performance of teams through selections of players, training of players, and implementation of tactics including formations  \cite{Dobson2011}.
In particular, a formation is a part of tactics to determine how players participate in offense and defense \cite{Bangsbo2000} and considered to affect match results  \cite{Bangsbo2000, Hirotsu2006}.
Managerial decision making in substituting players during a match may affect the probability of winning \cite{Hirotsu2006}.
We hypothesize that a manager continues to use the same formation if he has won the previous match, whereas he experiments on another formation following a loss in the previous match.

The WSLS and more general reinforcement learning posit that unsuccessful individuals modify their behavior to increase the probability of winning.
Therefore, we are interested in whether a formation change improves the performance of a team.
To clarify this point, we also investigate effects of formation changes on the results of succeeding matches.
 
\section{Materials and Methods}
\subsection{Data set}
We collected data on football matches from two websites, J-League Data Site (officially, ``J. League'') \cite{JLeague} on J-League, and Kicker-online \cite{Kicker} on Bundesliga.
J-League and Bundesliga are the most prestigious professional football leagues in Japan and Germany, respectively.
We refer to the two data sets as the J-League and the Bundesliga data sets.
The two data sets contain, for each team and match, the season, date, manager's name, result (i.e., win, draw, or loss), and starting formation.
Basic statistics of the data sets are summarized in Table \ref{table:summary1}.
The distributions of the probability of winning for teams and managers are shown in Fig. 1 for the two data sets.
Because the strength of a team apart from the manager was considered to affect the probability of winning, in Fig. 1, we treated a manager as different data points when he directed different teams.
The same caveat applies to all the following analysis focusing on individual managers (Figs. 3--6).

Between 1993 and 2004, except for 1996, each season of J-League was divided into two half seasons. 
After the two half seasons had been completed, two champion teams, each representing a half season, played play-off matches.
We regarded each half season as a season because intervals between two half seasons ranged from ten days to two months and therefore are longer than one week, which was a typical interval between two matches within a season.
We also carried out the same analysis when we regarded one year, not one half season, as a season and verified that the main results were unaltered (Appendix A).

We also collected data on Bundesliga from another website, Fussballdaten \cite{Fussballdaten}.
We focused on the Kicker-online data rather than the Fullballdaten data because the definition of the position was coarser for the Fussballdaten data (i.e., a player was not assumed to change his position during a season) than the Kicker-online data. 
Nevertheless, to verify the robustness of the following results, we also analyzed the Fussballdaten data (Appendix B).

\subsection{Definition of formation}
The definition of formation was different between the two data sets.
In the J-League data, each of the ten field players was assigned to either defender (DF), midfielder (MF), or forward (FW) in each match.
We defined formation as a triplet of the numbers of DF, MF, and FW players, which sum up to ten.
For example, a formation 4-4-2 implies four DFs, four MFs, and two FWs.

In the Bundesliga data, the starting positions of the players were given on a two-dimensional map of the pitch (Fig. \ref{kicker}).
For this data set, we defined formation as follows.
First, we measured the distance between the goal line and the bottom edge of the image representing each player along the vertical axis  (e.g., 113 shown in Fig. 2). 
We referred to the HTML source code of Kicker-online to do this. 
The unit of the distance is pixel (px).
The distance between the goal line and the half-way line is between 45 and 60 m in real fields. 
The same distance is approximately equal to 500 px in Kicker-online. Therefore, 1 px in Kicker-online roughly corresponds to 10 cm in real fields. 
Although the HTML source code also included the distance of players from the left touch line, we neglected this information because the primary determinant of the player's position seems to be the distance from the goal line rather than that from the left or right touch line, as implied by the terms DF, MF, and FW.
Second, we grouped players whose distances from the goal line were the same.
Third, we ordered the groups of players in terms of the distance, resulting in an ordered set of the numbers of players at each distance value.
The set of numbers defined a formation.
For example, when the distances of the ten field players are equal to 113 px, 113 px, 113 px, 113 px, 236 px, 236 px, 359 px, 359 px, 359 px, and 441 px, the formation of the team is defined to be 4-2-3-1 (Fig. \ref{kicker}). 

Among all matches in the Bundesliga data, the smallest nonzero distance between two players was equal to 31 px. 
Therefore, we did not have to worry about the possibility that players possessed almost the same distance values while being classified into distinct positions. 
For example, there was no case in which the distances of two field players from the goal line were equal to 113 px and 114 px.

For both data sets, a formation was defined as an ordered set of numbers, whereas the definition differs for the two data sets. 
For example, forward players possessing distance values 359 and 441 were classified into different positions in the Bundesliga data, whereas they belonged to the same position in the J-League data if they were both assigned to FW. 
In the following, we regarded that formation was changed when the ordered set of numbers differed between two consecutive matches.

Figures \ref{hist_fchange}(a)--(d) show the distribution of the probability that a team or manager has changed the formation in the two leagues.
To calculate the probability of formation changes for a team, we excluded the first match in each season and the matches immediately after a change in the manager.
As in the case of formation changes, we regarded that a manager was changed when the manager directed a team in a given match but did not do so in the next match. 
With this definition, a short absence of a chief manager due to illness, for example, may induce formation changes. 
However, we adhered to this definition because of the lack of further information on behavior of managers.
In addition, as explained in section 2.1, we treated a manager as different data points when he led different teams.

The frequency of formation changes as a function of time is shown in Figs. 3(e) and 3(f) for J-League and Bundesliga, respectively. 
The figures suggest that the frequency of formation change is stable over years in J-League, but not in Bundesliga.
Finally, we measured burstiness and memory coefficient \cite{Goh2008} for interevent times of formation changes to quantify temporal patterns of formation changes. 
The results are shown in Appendix C.

\subsection{GLMM}
To statistically examine whether patterns of formation changes were consistent with WSLS behavior, we investigated effects of previous matches and other factors on the likelihood of formation change for each team. 
If managers used the WSLS, the effect of the win and loss in the previous match on the likelihood of formation change should be significantly negative and positive, respectively.
We used a generalized linear mixed model (GLMM) with binomial errors and a logit-link function.

The dependent variable was the occurrence or lack thereof of formation changes, which was binary.
As independent variables, we included the binary variable representing whether or not the stadium was the home of the team (i.e., home or away) and the ternary result of the previous match (i.e., win, draw, or loss).
We designated the draw as the reference category for the match result.
Because the likelihood of formation changes may be affected by a streak of wins or losses, we also included the result of the second last match as an independent variable.
The difference between the focal team's strength and the opponent's strength was also an independent variable.
The strength of a team was defined by the probability of winning in the season.
We estimated the strength of a team separately for each season because it can vary across seasons.
The name of the manager was included as a random effect (random intercept). 

In this and the following analysis, we excluded the first match in each season for each team because we considered that the result of the last match in the preceding season would not directly affect the first match in a new season.
In addition, we excluded matches immediately after a change of manager because we were not interested in formation changes induced by a change of manager.
We further excluded the second match in each season for each team from the GLMM analysis when we employed the result of the second last match as an independent variable.
Because the J-League data set did not have the information on managers between 1993 and 1998, we only used data between 1999 and 2014 in the GLMM analysis.
We performed the statistical analysis using R 3.1.2 \cite{R2014} with lme4 package \cite{lme4}.

\subsection{Ordered probit model}
We also investigated the effects of formation changes on match results.
We used the ordered probit model because a match result was ternary.
Because the strength was considered to heavily depend on teams, we controlled for the strength of teams.
The same model was used for fitting match results in football in the Netherlands \cite{Koning2000} and the UK \cite{Dobson2011}.

The dependent variable of the model was a match result.
We assumed that the occurrence of formation change (change or no change), the stadium (home or away), the strength of teams, and the result of the previous match (win, draw, or loss) can affect a match result.
As a linear combination of these factors, we defined the unobserved potential variable for team $i$, denoted by $\alpha_i$, by
\begin{equation}
\alpha_{i} = \beta_{\rm f}f_{i} + \beta_{\rm h}h_{i}+\beta_{\rm w}w_{i} + \beta_{\rm \ell}\ell_{i} +\beta_{\rm r}r_{i},
\end{equation}
where $f_{i}=1$ if team $i$ changed the formation, and $f_{i}=0$ otherwise;
$h_{i}=1$ if the stadium was the home of team $i$, and $h_{i}=0$ otherwise;
$w_{i} = 1$ if team $i$ won the previous match, and $w_{i}=0$ otherwise;
$\ell_{i} = 1$ if team $i$ lost the previous match, and $\ell_{i}=0$ otherwise;
the strength of team $i$ denoted by $r_{i}$ was defined as the fraction of matches that team $i$ won in the given season.
In Appendix D, we conducted the analysis by assuming that $r_{i}$ was a latent variable obeying the normal distribution and then using the hierarchical Bayesian model  \cite{Parent2012}.

Consider a match between home team $i$ and away team $j$.
We assumed that the match result, denoted by $k_{ij}$, was determined by the difference between the potential values of the two teams, i.e., 

\begin{equation}
y_{ij} \equiv \alpha_{i} - \alpha_{j}.
\label{eq:y}
\end{equation}

Variables $y_{ij}$ and $k_{ij}$ are related by

\begin{equation}
\label{eq:probit}
  k_{ij} = \begin{cases}
    2 \mbox{ (home team wins) } & \mbox{if }\:\:  c_{1}<y_{ij} + \epsilon_{ij}, \\
    1 \mbox{ (draw) } & \mbox{if }\:\: c_{0}<y_{ij} + \epsilon_{ij} <c_{1}, \\
    0 \mbox{ (home team loses) } & \mbox{if }\:\: y_{ij} + \epsilon_{ij}<c_{0},
  \end{cases}
\end{equation}
where $c_0$ and $c_{1}$ are threshold parameters, and $\epsilon_{ij}$ is an error term that obeys the normal distribution with mean $0$ and standard deviation $1$.
Because $h_{i}-h_{j}=1$, $\beta_{\rm h}$ appears as a constant term on the right-hand side of Eq.~\eqref{eq:y}.
In fact, it is impossible to estimate $\beta_{\rm h}$ because $\beta_{\rm h}$ effectively shifts $c_0$ and $c_1$ by the same amount such that there are only two degrees of freedom in the parameter space spanned by $c_0$, $c_1$, and $\beta_{\rm h}$.
Therefore, we assumed $c_{0}=-c_{1}$ and estimated $c_0$ and $\beta_{\rm h}$. 
This assumption did not alter the estimates of the other parameters.
Equation (\ref{eq:probit}) results in
\begin{eqnarray}
P(k_{ij}=2) &=& 1-\Phi(c_{1}-y_{ij}),\\
P(k_{ij}=1) &=& \Phi(c_{1}-y_{ij}) - \Phi(c_{0}-y_{ij}),\\
P(k_{ij}=0) &=& \Phi(c_{0}-y_{ij}),
\end{eqnarray}
where $P$ denotes the probability, and $\Phi(\cdot)$ is the cumulative standard normal distribution function.

We excluded the matches that were the first game in a season at least for either team.
We also excluded matches immediately after a change of manager in either team.
Because the J-League data set did not contain the information on managers between 1993 and 1998, we only used data between 1999 and 2014 in this analysis.
We performed the analysis using R 3.1.2 \cite{R2014} and maxLik package \cite{maxLik}.

\subsection{Influence of individual manager's behavior on match results}

Different managers may show WSLS behavior to different extents to respectively affect match results. 
Therefore, we analyzed data separately for individual managers.
For each manager $i$, we calculated the probability of winning under each of the following four conditions: (i) $i$'s team won the previous match, and $i$ changed the formation, (ii) $i$'s team won the previous match, and $i$ did not change the formation, (iii) $i$'s team lost the previous match, and $i$ changed the formation, and (iv) $i$'s team lost the previous match, and $i$ did not change the formation. 
We then compared the probability of winning between cases (i) and (ii), and between cases (iii) and (iv) using the paired t-test.
In the t-test, we included the managers who directed at least ten pairs of consecutive matches in both of the two cases in comparison.
In this and the next sections, we treated a manager as different data points when he directed different teams, as explained in section 2.1.
In addition, we excluded the pairs of consecutive matches when the managers changed the team between the two matches.

\subsection{Degree of win-stay lose-shift}

To further examine possible relationships between manager's behavior and match results, we looked at the relationships between the tendency of the WSLS behavior for each manager (degree of WSLS for short) and the probability of winning. 
The degree of WSLS is defined by
\begin{eqnarray}
&&\mbox{degree of WSLS} \nonumber \\ 
&=&\bigl|P(\mbox{change}|\mbox{win}) - P_{\rm{WSLS}}(\mbox{change}|\mbox{win})\bigr|+\bigl|P(\mbox{change}|\mbox{loss})-P_{\rm{WSLS}}(\mbox{change}|\mbox{loss})\bigr| \nonumber\\
&=& \bigl|P(\mbox{change}|\mbox{win}) - 0\bigr|+\bigl|P(\mbox{change}|\mbox{loss}) - 1\bigr| \nonumber\\
&=& P(\mbox{change}|\mbox{win})+1-P(\mbox{change}|\mbox{loss}),
\end{eqnarray}
where $P_{\rm{WSLS}}(\mbox{change}|\mbox{win})$ ($=0$) is the conditional probability that a perfect WSLS manager changes the formation after winning, and likewise for $P_{\rm{WSLS}}(\mbox{change}|\mbox{loss})$ ($=1$). 
The degree of WSLS ranges from $0$ to $2$.\\

\section{Results}

\subsection{Win-stay lose-shift behavior in formation changes}
We examined the extent to which managers possibly changed the formation of the team after losing a match and persist to the current formation after a win.
The results of the GLMM analysis with the results of the previous matches being the only independent variables are shown in Table \ref{table:glmm1}.
For both data sets, winning in a match significantly decreased the probability of formation change in the next match, and losing in a match increased the probability of formation change.
The results did not essentially change when we used the full set of independent variables (Table \ref{table:glmm2}).
Formation changes are consistent with WSLS patterns.

For the J-League data, the effects of all the additional independent variables were insignificant.
We analyzed the J-League data by regarding a pair of half seasons (i.e., an yearly season) as a season to confirm that the results remained qualitatively the same except that winning in the second last match also significantly decreased the probability of formation change (Appendix A).
We also confirmed that matches played in the further past affect the probability of formation change to progressively small extents (Appendix E).

For the Bundesliga data, winning in the second last match also significantly decreased the probability of formation change in the extended GLMM model (Table \ref{table:glmm2}). 
These results are consistent with WSLS behavior. 
We also found for the Bundesliga data that stronger teams less frequently changed the formation and that a team would not change the formation to fight home games.
We also investigated the Fussballdaten data for Bundesliga, in which the definition of formation was different, and confirmed that managers tended to use the WSLS strategy (Appendix B).

\subsection{Determinants of match results}
The results obtained from the ordered probit model are shown in Table \ref{table:probit}.
For both data sets, formation changes did not significantly affect a match result.
The result remained qualitatively the same when each pair of half seasons was considered as a season in the J-League data (Appendix A), and when the strength of team was assumed to be a latent variable in the ordered probit model (Appendix D).
However, when the Fussballdaten data were used, formation changes significantly decreased the probability of winning (Appendix B).
Table \ref{table:probit} also tells us the following. 
Trivially, stronger teams were more likely to win in both data sets.
The home advantage was significant in both data sets, consistent with previous literature \cite{Albert2010, Dobson2011}.
In Bundesliga, a win tended to yield a poor result in the next match.
This is consistent with negative persistence effects reported in previous literature \cite{Dobson2011}, i.e., the results of the current and previous matches tend to be the opposite.

Figure \ref{winrate}(a) shows the probability of winning after individual managers changed or did not change the formation after a win in the J-League data.
A large circle in Figure \ref{winrate} represents a manager who presented both types of actions (i.e., formation change after winning and no formation change after winning) at least ten times. 
A small circle represents a manager who presented either type of action less than ten times.
The formation change does not appear to affect the probability of winning. 
This is also apparently the case for the actions after a loss (Figure  \ref{winrate}(b)) and the Bundesliga data (Figures  \ref{winrate}(c) and  \ref{winrate}(d)).
The results also appear to be insensitive to the unconditional probability of winning, which roughly corresponds to the position along the diagonal in Figure 4.
To be quantitative, we conducted the paired t-test on the managers who submitted the two types of actions at least ten times in each case (managers shown by the large circles in Figure  \ref{winrate}).
For the J-League data, there was no significant effect of formation change on the probability of winning both after winning ($p = 0.441$, $n=3$; corresponding to Figure  \ref{winrate}(a)) and losing ($p = 0.404$, $n=4$; Figure  \ref{winrate}(b)).
For the Bundesliga data, formation changes after winning significantly decreased the probability of winning in the next match ($p = 0.026$, $n=46$; Figure  \ref{winrate}(c)), whereas there was no significant effect after losing  ($p = 0.533$, $n=42$; Figure  \ref{winrate}(d)).
These results suggest that formation changes did not at least increase the possibility of winning.

The analysis with the ordered probit model aggregated the data from all managers. 
Therefore, we examined the relationship between the degree of WSLS and the probability of winning for individual managers.
The results are shown in Figure \ref{performance}. 
A circle in Figure \ref{performance} represents a manager.
We did not find a significant relationship between the usage of the WSLS and the probability of winning for both J-League (Pearson's $r = 0.213$, $p = 0.411$, $n=17$) and Bundesliga ($r = 0.058$, $p =  0.668$, $n=58$) data.

\section{Discussion}
We have provided evidence that football managers tend to stick to the current formation until the team loses, consistent with the WSLS strategy previously shown in laboratory experiments with social dilemma games \cite{Wedekind1996, Milinski1998} and gambling tasks \cite{Hayden2009, Scheibehenne2011}. 
Formation changes did not significantly affect (at least did not improve) a match result in most cases.
This result seems to be odd because managers change formation to lead the team to a success.
Generally speaking, when the environment in which an agent is located is fixed or exogenously changing, reinforcement learning usually improves the performance of the agent \cite{Sutton1998}. 
However, computational studies have suggested that it is not always the case when agents employing reinforcement learning are competing with each other,  because the competing agents try to supersede each other \cite{Taiji1999, Macy2002, Masuda2009, Masuda2011}.
The present finding that manager's WSLS behavior does not improve team's performance is consistent with these computational results.  
Empirical studies also suggest that humans obeying reinforcement learning does not improve the performance in complex environments.
For example, players in the National Basketball Association were more likely to attempt 3 point shots after successful 3 point shots although their probability of success decreased for additional shots \cite{Neiman2011}.
Also in nonscientific accounts, it has been suggested that humans engaged in sports and gambles often use the WSLS strategy even if outcome of games is determined merely at random \cite{Vyse2013}.
We have provided quantitative evidence underlying these statements. 

Many sports fans possess the hot hand belief in match results, i.e., belief that a win or good performance persists \cite{Bar-Eli2006}. 
However, empirical evidence supports that streaks of wins and those of losses are less likely to occur than under the independence assumption \cite{Bar-Eli2006}.
By analyzing patterns of matches in the top division of football in England, Dobson and Goddard suggested the existence of negative persistence effects, i.e., a team with consecutive wins tended to perform poorly in the next match and vice versa \cite{Dobson2011}. 
Their results are consistent with the present results; we observed the negative persistence effects, i.e., anticorrelation between the results of the previous and present matches. 

In the present study, we have neglected various factors that potentially affect the likelihood of formation change because our data sets did not contain the relevant information.
For example, managers may change formations due to injuries, suspensions of players, and other strategic reasons including transfer of players.
More detailed data will be able to provide further understanding of the relative importance of strategic versus accidental factors in formation changes.

An important limitation of the present study is that we have oversimplified the concept of formation.
Effective formations dynamically change during a match owing to movements of players.
Because of the availability of data and our interests in the manager's long-term behavior rather than formation changes during a match \cite{Hirotsu2006}, we used the formation data released in the beginning of the matches.
Based on recent technological developments, formations can be extracted from tracking data on movement patterns of players \cite{Bialkowski2014a, Bialkowski2014b}.
Investigations on manager's decision making using such technologies warrant further research.

\section*{Appendix A: Analysis of the J-League data on the basis of yearly seasons}
From 1993 to 2004, except for 1996, each season of J-League, spanning a year, was subdivided into two half seasons.
In the main text, we regarded each half season as a season.
To examine the robustness of our results, we carried out the same analysis when we regarded one entire season (i.e., one year), not one half season, as a season.
The results were qualitatively the same as those shown in the main text (Tables \ref{table:j_half_glmm1}--\ref{table:j_half_probit}) except that winning in the second last match significantly decreased the probability of formation change.

\section*{Appendix B: Fussballdaten}
We analyzed data on Bundesliga from another website, Fussballdaten \cite{Fussballdaten}.
In the Fussballdaten data, each field player was assigned to one of the three positions (i.e., DF, MF, or FW) registered for an entire season.
We defined the formation by counting the number of each type of field player in the same manner as that for the J-League data set.

First, to examine possible existence of WSLS behavior by managers, we applied the GLMM analysis to the Fussballdaten data.
The results shown in Tables \ref{table:fussball_glmm1} and \ref{table:fussball_glmm2} are largely consistent with those for the Kicker online data (Tables 1 and 2).
In particular, winning and losing in the previous match significantly decreased and increased the probability of formation change in the next match, respectively, consistent with WSLS behavior.

Second, we also investigated the effect of formation change and other factors on the match result using the ordered probit model.
Table \ref{table:fussball_probit} indicates that formation changes have decreased the probability of winning.
This result is not consistent with those for the two data sets shown in the main text.
In addition, winning in the previous match decreased the probability of winning in the next match, indicating the presence of the negative persistence effect.
This result is consistent with that for the Kicker online data (Table \ref{table:probit}).

\section*{Appendix C: Burstiness and memory coefficient of interevent time series} 

To capture temporal properties of formation changes, in this section we calculated burstiness, $B$, and memory coefficient, $M$, \cite{Goh2008} on the basis of the interevent time series $\{ \tau_i \}$ defined as follows.
We calculated $B$ and $M$ for each manager. As in the main text, we treated a manager as different data points when he directed different teams.
In the main text, we used interevent time series for individual seasons without concatenating different seasons.
In this section, however, we use $\{\tau^{k}_i\}$ obtained by concatenating all seasons.
For a given manager, we denote by $t_0$, $t_1$, $\ldots$, $t_{N}$ ($2\le t_0 < t_1 < \cdots < t_N$) the times when the manager changes the formation. 
The number of formation change summed over all seasons is equal to $N+1$.
We counted the time in terms of the number of match rather than the day to exclude effects of variable intervals between consecutive matches in terms of the real time.
If $t_2 = 5$, for example, the manager changed the formation to play the fifth match, and it was the third change for the manager since the first match in the data set.

Managers sometimes moved from one team to another or did not direct any team.
Because formation changes occurring as a result of a manager's move or after a long absence were not considered to be strategic, we discarded the corresponding intervals.
It should be noted that we did not mix interevent time series for a manager leading different teams.
Then, a time series $\{t_i\}$ for each manager was partitioned into $N_k$ segments by either a manager's move or absence.
We denoted by $N_k+1$ and $K$ the total number of formation changes in the $k$th segment and the total number of segments, respectively. 
It holds true that $\sum_{k=1}^{K} (N_k + 1) = N+1$.
We also denoted by $t^{k}_{i}$ $(0\le i\le N_k)$ the time of the $i$th formation change in the $k$th segment.
The interevent time for formation changes was defined by $\tau^{k}_{i} = t^{k}_{i}-t^{k}_{i-1}$ $(1\le i\le N_k)$. 
It holds true that $\sum_{k=1}^{K} \sum_{i=1}^{N_k}\tau^{k}_{i}=N+1-K$.
In the following analysis, we used managers who directed at least 100 matches in a team and has $N+1-K \ge 10$.

The burstiness is defined by
\begin{equation}
B=\frac{\sigma/m-1}{\sigma/m+1}=\frac{\sigma-m}{\sigma+m},
\end{equation}
where $m=\sum_{k=1}^{K} \sum_{i=1}^{N_{k}} \tau^{k}_{i}/(N+1-K)$ and $\sigma=\sqrt{ \sum_{k=1}^{K} \sum_{i=1}^{N_{k}} (\tau^{k}_{i} - m)^2 /(N+1-K)}$ are the mean and standard deviation of interevent time, respectively.
$B$ ranges between $-1$ and $1$.
A large value of $B$ indicates that a sequence of formation change events is bursty in the sense that the interevent time obeys a long-tailed distribution.
The Poisson process yields the exponential distribution and hence yields $B = 0$.

The memory coefficient quantifies the correlation between two consecutive interevent times and is defined by
\begin{equation}
M=\frac{1}{N-K} \sum_{k=1}^{K} \sum_{i=1}^{N_{k}-1}\frac{(\tau_{i}^{k}-m_{1})(\tau_{i+1}^{k}-m_{2})}{\sigma_{1}\sigma_{2}},
\end{equation} 
where 
\begin{equation}
m_{1}= \sum_{k=1}^{K} \sum_{i=1}^{N_{k}-1} \frac{\tau^{k}_i}{N-K},
\end{equation}
\begin{equation}
m_{2}= \sum_{k=1}^{K} \sum_{i=2}^{N_{k}}\frac{ \tau^{k}_i}{N-K},
\end{equation}
\begin{equation}
\sigma_{1}=\sqrt{\sum_{k=1}^{K} \sum_{i=1}^{N_{k}-1} \frac{(\tau^{k}_i - m_{1})^2}{N-K}},
\end{equation}
and 
\begin{equation}
\sigma_{2}=\sqrt{\sum_{k=1}^{K}\sum_{i=2}^{N_{k}} \frac{(\tau^{k}_i - m_{2})^2}{N-K}}.
\end{equation}
An uncorrelated sequence of interevent times yields $M=0$.

To examine the statistical significance of $B$ value for each manager, we generated $10^3$ sequences of interevent times from the exponential distribution whose mean was equal to that of the original data.
Each synthesized sequence had the same length (i.e., $N$) as that of the original data.
We calculated $B$ for each synthesized sequence.
We regarded that the value of $B$ for the original data was significant if it was not included in the 95\% confidential interval (CI) on the basis of the distribution generated by the $10^3$ sequences corresponding to the Poisson process.
We calculated the CI for $M$ in the same manner except that we generated synthesized sequences by randomizing the original sequence of interevent times, instead of sampling sequences from the exponential distribution.

Figure \ref{burst} shows histograms of burstiness, $B$, and memory coefficient, $M$, for the managers.
The average values of $B$ and $M$ for interevent times of formation changes in the J-League data were equal to $0.145$ and $-0.137$, respectively.
Those for the Bundesliga data set were equal to $0.022$ and $-0.120$, respectively.
For both data sets, the average values of $B$ were positive, and those of $M$ were negative.
The fraction of managers yielding significantly positive and negative $B$ values were equal to $0.385$ and $0$, respectively, for the J-League data.
Those for the Bundesliga data were equal to $0.372$ and $0.244$, respectively.
These results suggest that in both data sets, a moderate fraction of managers changed formations in a bursty manner.
In the Bundesliga data, however, some managers changed formations more regularly than expected from the Poisson process.
The fractions of significantly positive and negative $M$ values were equal to $0.077$ and $0.077$, respectively, for the J-League data.
Those for the Bundesliga data were equal to $0.026$ and $0.026$, respectively.
In both data sets, the fractions of managers with significant $M$ values were small, indicating that two consecutive interevent times were uncorrelated for a majority of managers.

\section*{Appendix D: Hierarchical Bayesian model}
In the main text, we used the fraction of matches that a team won in a season to define the strength of the team.
In this section, we analyze a model in which the strength of a team is assumed to be a latent variable.
We used the hierarchical Bayesian ordered probit model combined with the Markov Chain Monte Carlo (MCMC) method \cite{Parent2012}.
The model is the same as that used in the main text except for the derivation of the team strength.
We assumed that the prior of the strength of team $i$ in a season, denoted by $r_{i}$, obeyed the normal distribution with mean $0$ and variance $\sigma^2$.
The priors of $\beta_{\rm f}$, $\beta_{\rm h}$, $\beta_{\rm w}$, and $\beta_{\rm \ell}$ obeyed the normal distribution with mean $0$ and variance $10^2$. 
The prior of $\sigma^2$ obeyed the uniform distribution on [0, $10^4$].
We conducted MCMC simulations for four independent chains starting from the same prior distributions.
The total iterate per chain was set to 25,000, and the first 5,000 iterates were discarded as transient.
The thinning interval was set to 20 iterates.
A final coefficient was regarded to be significant if the 95\% credible interval did not bracket zero. 
We excluded the matches that were the first game in a season at least for either team.
We performed the analysis using R 3.1.2 \cite{R2014} and RStan package \cite{RStan2014}.

Table \ref{table:bayes} summarizes the results obtained from the Bayesian probit model.
For both data sets, the credible interval of the coefficient representing the effect of the formation change brackets zero. 
Therefore, we conclude that formation changes have not affected the probability of winning.

\section*{Appendix E: Cross-correlation analysis}
To further investigate possible relationships between formation changes and match results, we measured the cross-correlation between the two.
In this analysis, we did not exclude the first match in each season.
We used the teams that played at least 100 matches.
We set $f_{i,t}=1$ if team $i$ changes the formation in the $t$th match ($2\le t\le T_i$), where $T_{i}$ is the number of matches played by team $i$, and $f_{i,t}=0$ otherwise;
$w_{i,t} = 1$ if team $i$ wins in the $t$th match, and $w_{i,t}=0$ otherwise;
$\ell_{i,t} = 1$ if team $i$ loses in the $t$th match, and $\ell_{i,t}=0$ otherwise.
We defined the cross-correlation between two time series $\{x_{i,t}\}$ and $\{y_{i,t}\}$ by
\begin{equation}
 \rho(x,y,\tilde{\tau}) = \frac{\Sigma_{i=1}^{N_{\rm team}}\Sigma_{t=2}^{T_{i}-\tilde{\tau}}  (x_{i,t+\tilde{\tau}}-\bar{x})(y_{i,t}-\bar{y})}{\sqrt{\Sigma_{i=1}^{N_{\rm team}}\Sigma_{t=2}^{T_{i}-\tilde{\tau}} (x_{i,t+\tilde{\tau}}-\bar{x})^2}\sqrt{\Sigma_{i=1}^{N_{\rm team}}\Sigma_{t=2}^{T_{i}-\tilde{\tau}} (y_{i,t}-\bar{y})^2}},
\end{equation}
where $\bar{x}=(1/N_{\rm team})\times\Sigma_{i=1}^{N_{\rm team}}\Sigma_{t=2}^{T_{i}-\tilde{\tau}}x_{i,t+\tilde{\tau}}/(T_{i}-\tilde{\tau}-1)$, $\bar{y}=(1/N_{\rm team})\times\Sigma_{i=1}^{N_{\rm team}}\Sigma_{t=2}^{T_{i}-\tilde{\tau}}y_{i,t}/(T_{i}-\tilde{\tau}-1)$, $N_{\rm team}$ is the number of teams, and $\tilde{\tau}$ is the lag.
We measured the cross-correlation between formation changes and wins by
\begin{eqnarray}
\rho(f,w,\tilde{\tau}) & \mbox{if }\:\:  \tilde{\tau} \geq 0, \label{rho1} \\
\rho(w,f,-\tilde{\tau}) & \mbox{if }\:\:  \tilde{\tau} < 0. \label{rho2}
\end{eqnarray}
Replacing $w$ by $\ell$ in Eqs.~\eqref{rho1} and \eqref{rho2} defines the cross-correlation between formation changes and losses.

To examine the statistical significance of the cross-correlation obtained from the original data, we generated $10^3$ randomized sequences of formation changes as follows.
For given team $i$ and positive lag $\tilde{\tau}$, we randomly shuffled the original sequence of formation changes, $\{f_{i,2+\tilde{\tau}}, \ldots, f_{i,T_{i}} \}$, by assigning 1 (i.e., formation change) to each match with the equal probability such that the number of 1s in the synthesized sequence was the equal to that in the original sequence.
We generated a randomized sequence for each team. 
Then, we measured the cross-correlation between the randomized sequences of formation changes and $\{ w_{i,2}, \ldots, w_{i,T_i-\tilde{\tau}}\}$ or $\{ \ell_{i,2}, \ldots, \ell_{i,T_i-\tilde{\tau}}\}$ using Eq.~\eqref{rho1}. 
We repeated this procedure $10^3$ times to obtain $10^3$ cross-correlation values.
The cross-correlation for the original data was considered to be significant for a given $\tilde{\tau}$ if it was not included within the 95\% CI calculated on the basis of the $10^3$ correlation coefficient values for the randomized samples.
We also examined the statistical significance of the cross correlation for a negative lag on the basis of $10^3$ cross-correlation values between randomized sequences of $\{ f_{i,2}, \ldots, f_{i,T_i-\tilde{\tau}}\}$ and $\{w_{i,2+\tilde{\tau}}, \ldots, w_{i,T_{i}} \}$ or $\{\ell_{i,2+\tilde{\tau}}, \ldots, \ell_{i,T_{i}} \}$ using Eq.~\eqref{rho2}.

The cross-correlation measured for various lags is shown in Figure \ref{ccf}.
The cross-correlation value was the largest in the absolute value at $\tilde{\tau} = 1$.
The effect of past matches on formation changes was mildly significant between $\tilde{\tau} = 2$ and $\tilde{\tau} \approx 5$. 
The sign of the effect (i.e., positive or negative) was the same for different lag values, which complies with the concept of WSLS. 
When $\tilde{\tau} \leq 0$, the cross-correlation was insignificant or weakly significant even if $\tilde{\tau} \approx 0$. 
This result is suggestive of a causal relationship, i.e., a match result tends to cause a formation change.

\section*{Acknowledgements}
This work is supported by JST, CREST.

\newpage
\section*{Figures}
\begin{figure}[H]
\begin{center}
\includegraphics[width=150mm]{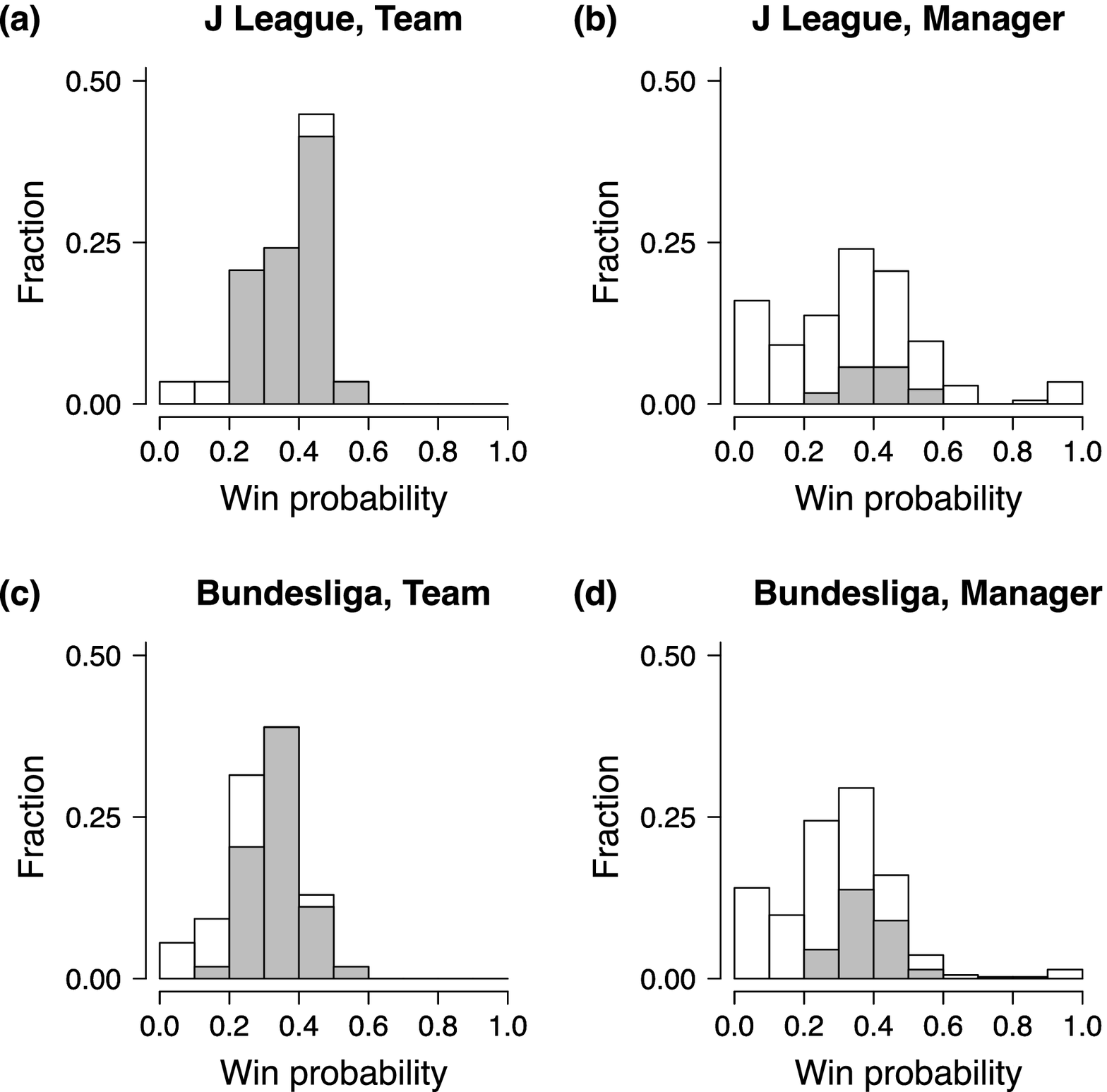}
\end{center}
\caption{Distributions of the probability of winning. (a) Distribution for the teams in the J-League data. (b) Distribution for the managers in the J-League data. (c) Distribution for the teams in the Bundesliga data. (d) Distribution for the managers in the Bundesliga data. The colored bars correspond to the teams or managers that have played at least 100 matches. }
\label{hist_win}
\end{figure}

\begin{figure}[H]
\begin{center}
\includegraphics[width=150mm]{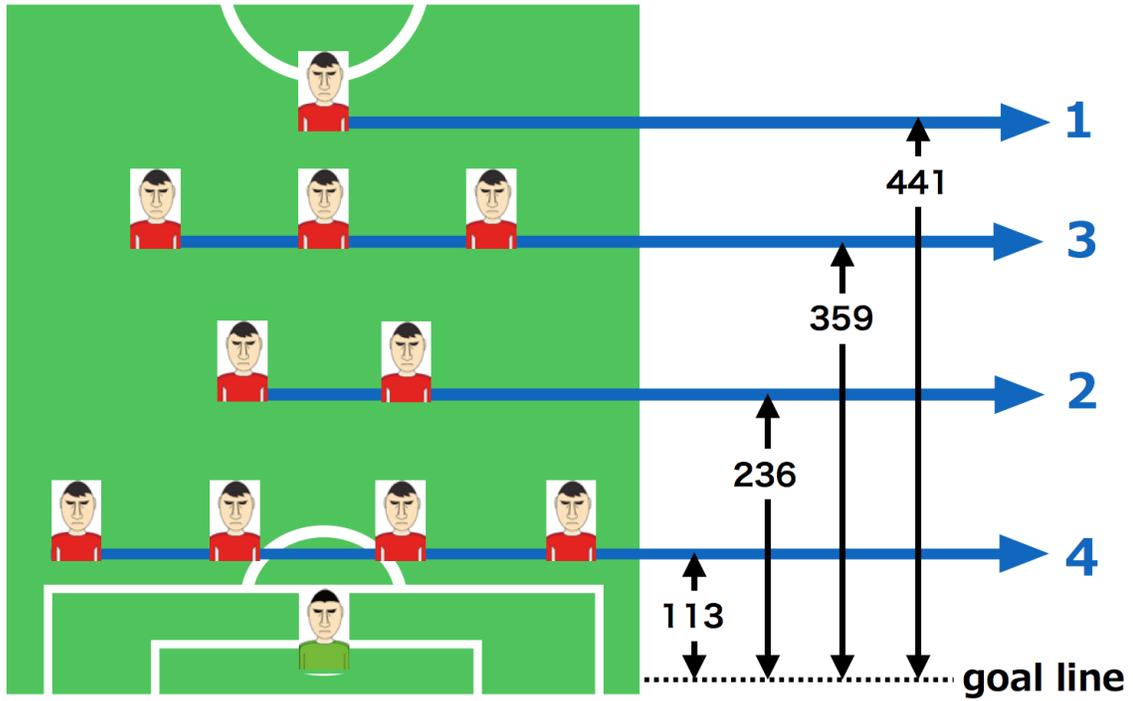}
\end{center}
\caption{Definition of formation in the Bundesliga data. Kicker online gives the starting positions of the players as two-dimensional coordinates on the pitch. Field players with the identical distance from the goal line are aggregated into the same position. The starting positions shown in the figure are coded as 4-2-3-1. }
\label{kicker}
\end{figure}

\begin{figure}[H]
\begin{center}
\includegraphics[width=110mm]{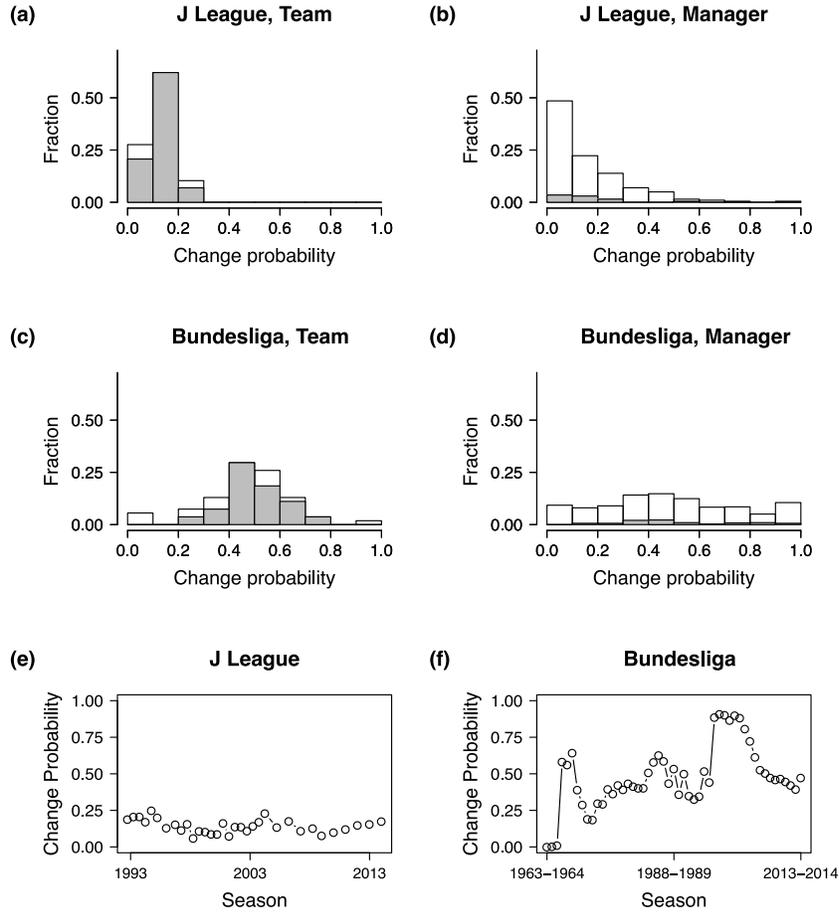}
\end{center}
\caption{Distributions of the probability of formation changes. (a) Distribution for the teams in the J-League data. (b) Distribution for the managers in the J-League data. (c) Distribution for the teams in the Bundesliga data. (d) Distribution for the managers in the Bundesliga data. In (a)--(d), the colored bars  correspond to the teams or managers that have played at least 100 matches. (e) Probability of formation change in each season in J-League, aggregated over the different teams and managers. A circle represents a season. Because the J-League data set did not have the information on managers between 1993 and 1998, we neglected changes of managers between 1993 and 1998. Between 1993 and 2004, except for 1996, intervals between two circles are dense because a season consists of two half seasons. (f) Probability of formation change in each season in Bundesliga, aggregated over the different teams and managers. }
\label{hist_fchange}
\end{figure}

\begin{figure}[H]
\begin{center}
\includegraphics[width=140mm]{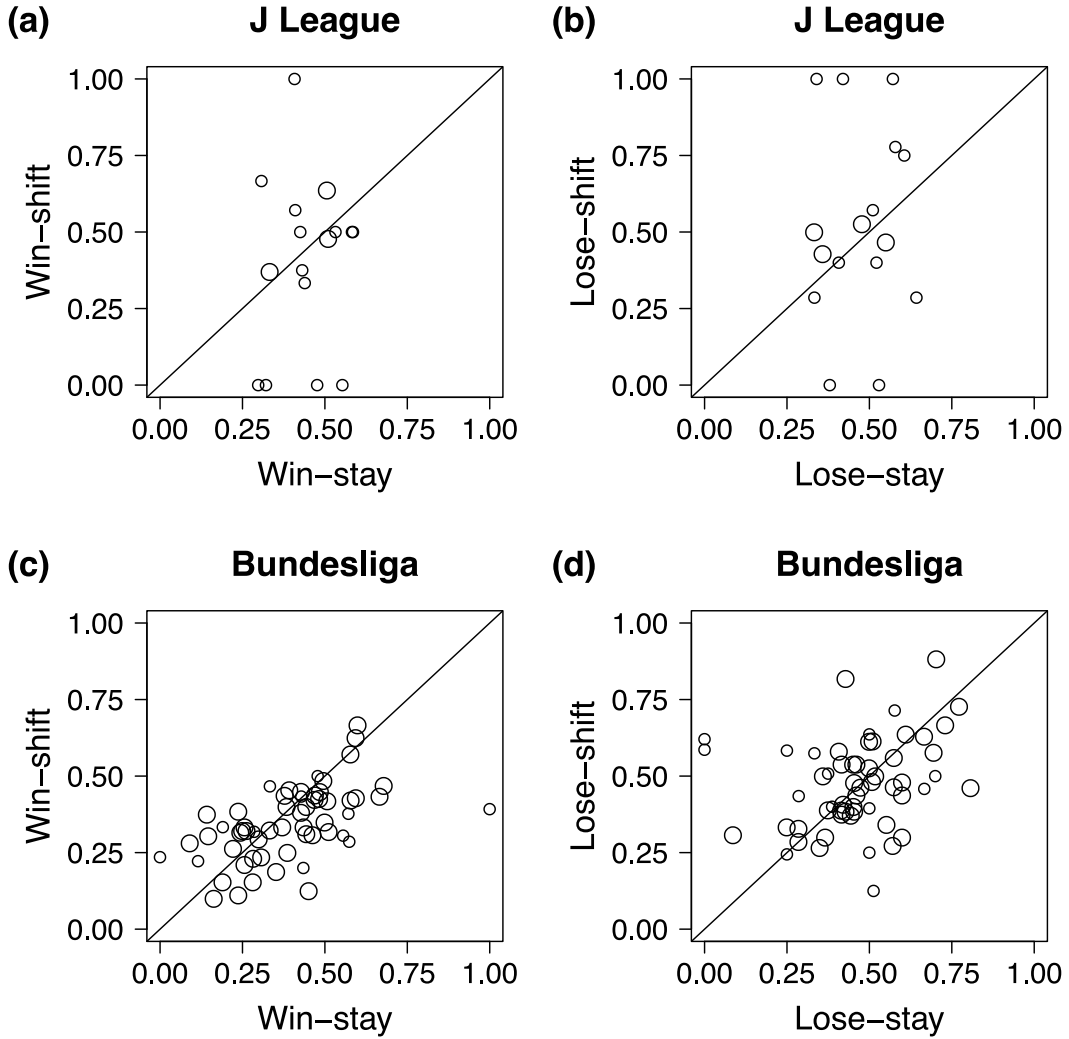}
\end{center}
\caption{Conditional probability of winning for individual managers after they changed or did not change the formation. A circle represents a manager associated with a team, who directed the team in at least 100 matches. A large circle represents a manager who showed both types of the actions (e.g., formation change after a win and no formation change after a win in (a) and (c)) at least ten times. A small circle represents a manager who presented either type of action less than ten times. (a) When the team won the previous match in the J-League data. (b) When the team lost the previous match in the J-League data. (c) When the team won the previous match in the Bundesliga data. (d) When the team lost the previous match in the Bundesliga data. }
\label{winrate}
\end{figure}

\begin{figure}[H]
\begin{center}
\includegraphics[width=150mm]{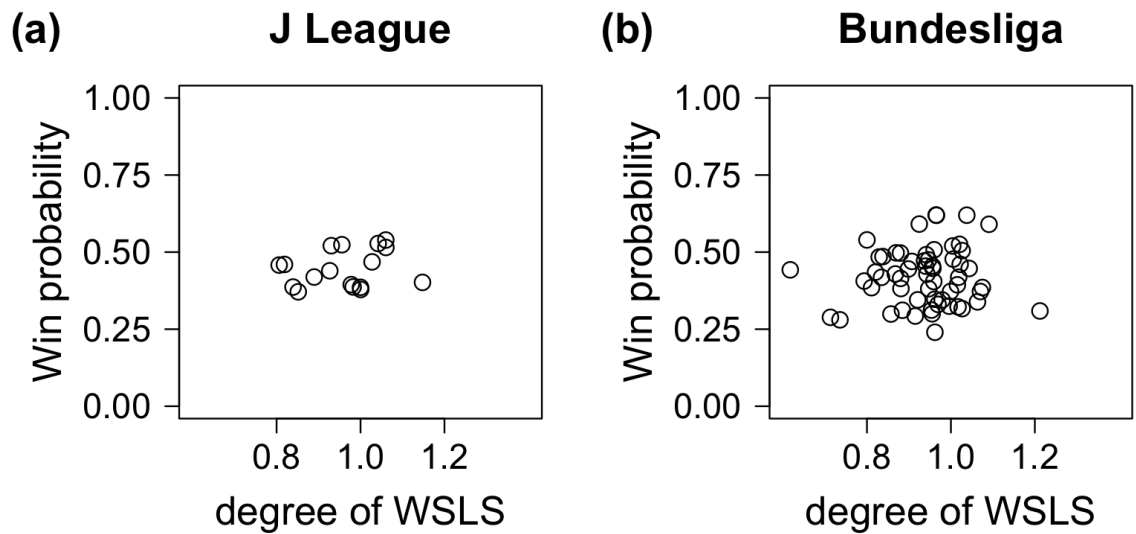}
\end{center}
\caption{Relationship between the degree of WSLS and the probability of winning. (a) J-League. (b) Bundesliga. A circle represents a manager associated with a team, who directed the team in at least 100 matches.}
\label{performance}
\end{figure}  

\begin{figure}[H]
\begin{center}
\includegraphics[width=150mm]{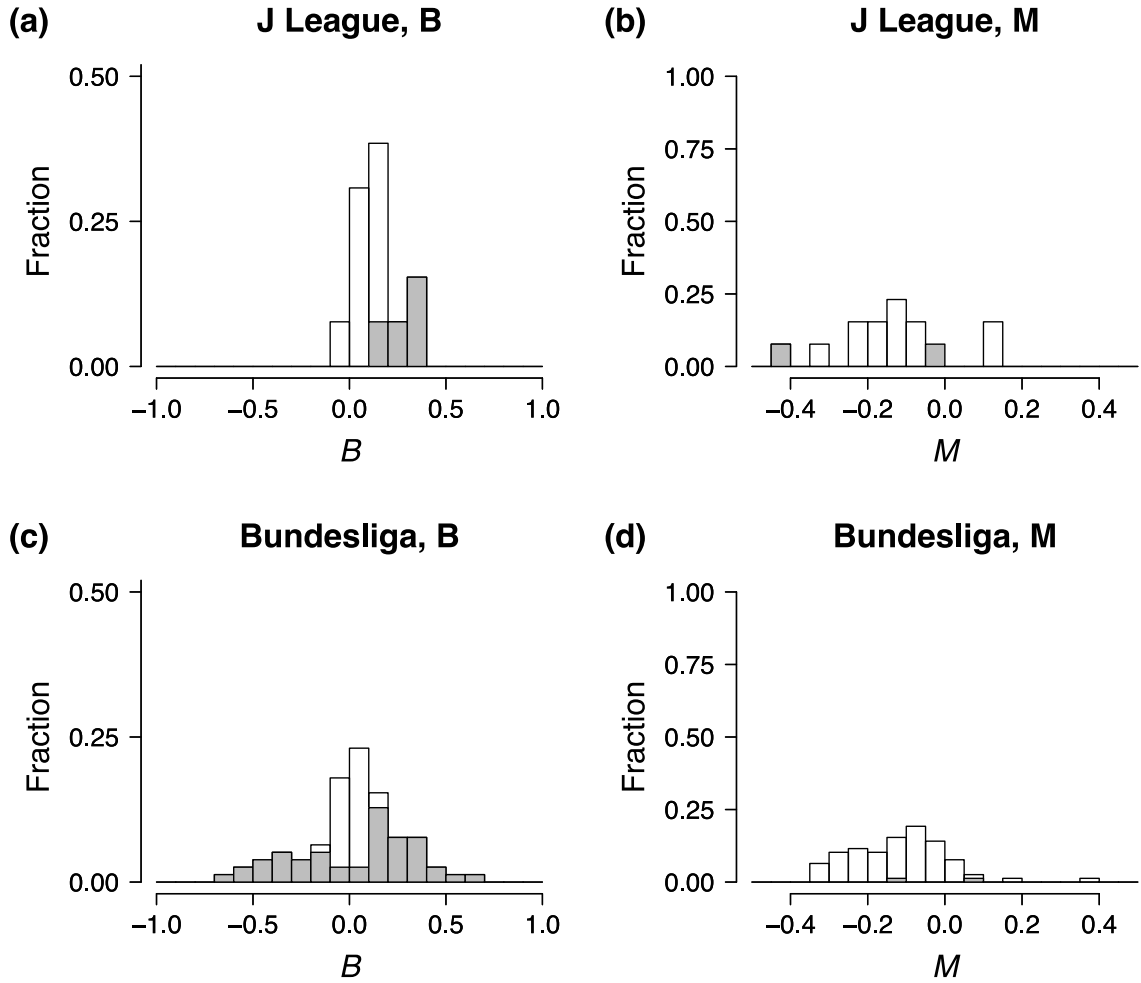}
\end{center}
\caption{Distributions of burstiness, $B$, and memory, $M$, across managers. (a) $B$ for the J-League data. (b) $M$ for the J-League data. (c) $B$ for the Bundesliga data. (d) $M$ for the Bundesliga data. The colored bars correspond to managers who have statistically significant values.}
\label{burst}
\end{figure}    

\begin{figure}[H]
\begin{center}
\includegraphics[width=150mm]{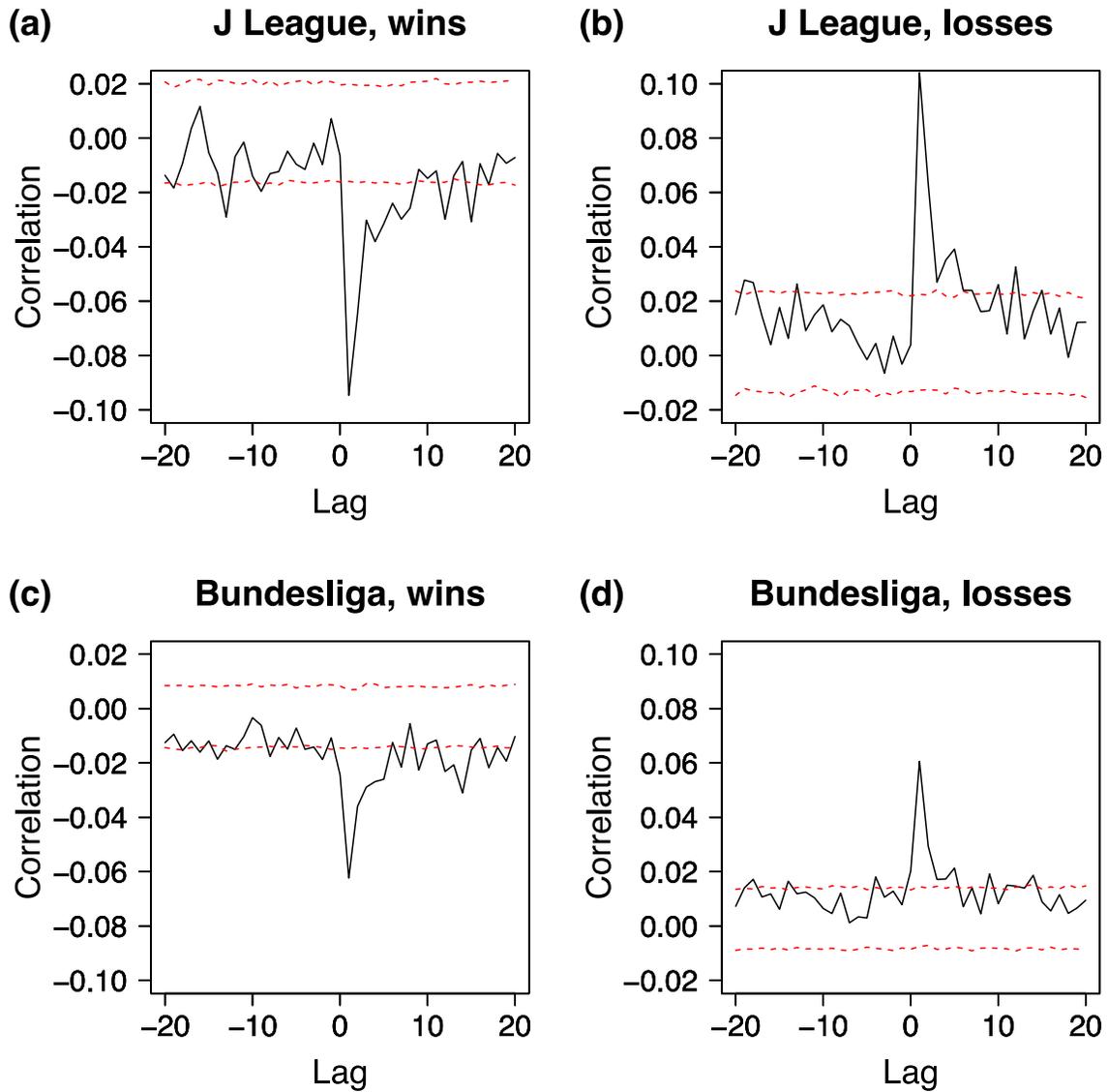}
\end{center}
\caption{Cross-correlation between temporal patterns of formation changes and match results. Ranges between the dashed lines represent 95\% CIs on the basis of the randomized sequences of formation changes. Cross-correlation between (a) formation changes and wins for the J-League data, (b) formation changes and losses for the J-League data, (c) formation changes and wins for the Bundesliga data, and (d) formation changes and losses for the Bundesliga data. }
\label{ccf}
\end{figure}

\newpage
\section*{Tables}
\begin{table}[H]
\begin{center}
  \begin{tabular}{ccccc}
    Quantities & J-League & Bundesliga \\ \hline
    Year & 1993--2014 (1999-2014) & 1963--2014 \\
    Number of seasons & 33 (22) & 51 \\
    Number of matches & 5,944 (4318) & 15,548 \\
    Number of teams & 29 (28) & 54 \\
    Number of managers & 176 & 372 \\
    Number of wins (or losses) & 4,961 (3,435) & 11,543 \\
    Number of draws & 983 (883) & 4,005 \\ \hline
  \end{tabular}
  \caption{Statistics of the J-League and Bundesliga data sets. Values for the subset of the J-League data for which the information about the managers is available are shown in the parentheses.}
  \label{table:summary1}
  \end{center}
\end{table}

\newpage
\begin{table}[H]
\begin{center}
  \begin{tabular}{ccccc}
    Data set & Variable & Coefficient & SE & p-value \\ \hline
    J-League & ${\rm Win}_{t-1}$ & $-0.360$ & $0.101$ & $< 0.001$ \\
    & ${\rm Loss}_{t-1}$ & $0.388$ & $0.093$ & $< 0.001$ \\  \hline
    Bundesliga & ${\rm Win}_{t-1}$ & $-0.179$ & $0.032$ & $< 0.001$ \\
    & ${\rm Loss}_{t-1}$ & $0.164$ & $0.032$ & $< 0.001$\\ \hline
  \end{tabular}
  \caption{Results of the GLMM analysis when the results of the previous match were used as the sole independent variables. ${\rm Win}_{t-1}$ denotes the binary variable representing whether the team has won the previous match ($0$: no win, $1$: win). Likewise for ${\rm Loss}_{t-1}$ ($0$: no loss, $1$: loss). SE: standard error. }
  \label{table:glmm1}
  \end{center}
\end{table}

\newpage
\begin{table}[H]
\begin{center}
  \begin{tabular}{ccccc}
    Data set & Variable & Coefficient & SE & p-value \\ \hline
    J-League & ${\rm Win}_{t-1}$ & $-0.299$ & $0.124$ & $0.016$ \\
    & ${\rm Win}_{t-2}$ & $-0.204$ & $0.116$ & $0.079$ \\
    & ${\rm Loss}_{t-1}$ & $0.387$ & $0.120$ & $0.001$ \\
    & ${\rm Loss}_{t-2}$ & $0.146$ & $0.126$ & $0.248$ \\
    & ${\rm Win}_{t-1}$ $\times$ ${\rm Win}_{t-2}$ & $0.005$ & $0.182$ & $0.979$ \\
    & ${\rm Loss}_{t-1}$ $\times$ ${\rm Loss}_{t-2}$ & $0.023$ & $0.164$ & $0.888$ \\
    & Home & $-0.062$ & $0.072$ & $0.392$ \\ 
    & Strength & $-0.192$ & $0.202$ & $0.343$ \\ \hline
    Bundesliga & ${\rm Win}_{t-1}$ & $-0.207$ & $0.040$ & $< 0.001$ \\
    & ${\rm Win}_{t-2}$ & $-0.117$ & $0.039$ & $0.003$ \\
    & ${\rm Loss}_{t-1}$ & $0.136$ & $0.039$ & $< 0.001$ \\
    & ${\rm Loss}_{t-2}$ & $0.007$ & $0.040$ & $0.867$ \\
    & ${\rm Win}_{t-1}$ $\times$ ${\rm Win}_{t-2}$ & $0.025$ & $0.059$ & $0.676$ \\
    &  ${\rm Loss}_{t-1}$ $\times$ ${\rm Loss}_{t-2}$ & $0.108$ & $0.060$ & $0.072$ \\
    & Home & $-0.118$ & $0.027$ & $< 0.001$ \\ 
    & Strength & $-0.530$ & $0.082$ & $< 0.001$ \\ \hline
  \end{tabular}
  \caption{Results of the GLMM analysis when all the independent variables were considered. ${\rm Win}_{t-i}$ ($i=1, 2$) is the binary variable representing whether or not the team has won the ($t-i$)th match ($0$: no win, $1$: win). Likewise for ${\rm Loss}_{t-i}$ ($i=1, 2$) (0: no loss, 1: loss). Home is equal to $0$ for an away game and $1$ for a home game. Strength is equal to the fraction of matches that the team has won in a season. SE: standard error.}
  \label{table:glmm2}
  \end{center}
\end{table}

\newpage
\begin{table}[H]
\begin{center}
  \begin{tabular}{ccccc}
    Data set & Variable & Coefficient & SE & p-value \\ \hline
    J-League & Formation change ($\beta_{\rm f}$) & $0.112$ & $0.063$ & $0.075$ \\
    & Home ($\beta_{\rm h}$) & $0.087$ & $0.033$ & $0.009$ \\
    & Win ($\beta_{\rm w}$) & $0.082$ & $0.099$ & $0.405$ \\
    & Loss ($\beta_{\rm \ell}$) & $0.193$ & $0.100$ & $0.053$ \\
    & Strength ($\beta_{\rm r}$) & $2.889$ & $0.169$ & $< 0.001$ \\ \hline
    Bundesliga & Formation change ($\beta_{\rm f}$) & $-0.011$ & $ 0.016$ & $0.509$ \\
    & Home ($\beta_{\rm h}$) & $0.420$ & $0.011$ & $< 0.001$ \\
    & Win ($\beta_{\rm w}$) & $-0.060$ & $0.018$ & $0.001$ \\
    & Loss ($\beta_{\rm \ell}$) & $-0.006$ & $0.018$ & $0.738$ \\
    & Strength ($\beta_{\rm r}$) & $2.792$ & $0.060$ & $< 0.001$ \\ \hline
  \end{tabular}
  \caption{Effects of variables on match results as obtained from the ordered probit model. SE: standard error.}
  \label{table:probit}
  \end{center}
\end{table}

\newpage
\begin{table}[H]
\begin{center}
  \begin{tabular}{cccc}
    Variable & Coefficient & SE & p-value \\ \hline
    ${\rm Win}_{t-1}$  & $-0.352$ & $0.101$ & $< 0.001$ \\
    ${\rm Loss}_{t-1}$  & $0.398$ & $0.093$ & $< 0.001$ \\  \hline
  \end{tabular}
  \caption{Results of the GLMM analysis for the J-League data when a year was regarded as a season. The results of the previous match were used as the sole independent variables. ${\rm Win}_{t-1}$ denotes the binary variable representing whether the team has won the previous match ($0$: no win, $1$: win). Likewise for ${\rm Loss}_{t-1}$ ($0$: no loss, $1$: loss). SE: standard error. }
  \label{table:j_half_glmm1}
  \end{center}
\end{table}

\newpage
\begin{table}[H]
\begin{center}
  \begin{tabular}{cccc}
    Variable & Coefficient & SE & p-value \\ \hline
    ${\rm Win}_{t-1}$ & $-0.306$ & $0.123$ & $0.013$ \\
    ${\rm Win}_{t-2}$ & $-0.223$ & $0.111$ & $0.045$ \\
    ${\rm Loss}_{t-1}$ & $0.406$ & $0.119$ & $< 0.001$ \\
    ${\rm Loss}_{t-2}$ & $0.145$ & $0.126$ & $0.249$ \\
    ${\rm Win}_{t-1}$ $\times$ ${\rm Win}_{t-2}$ & $0.021$ & $0.180$ & $0.909$ \\
    ${\rm Loss}_{t-1}$ $\times$ ${\rm Loss}_{t-2}$ & $ 0.003$ & $0.162$ & $0.985$ \\
    Home & $-0.062$ & $0.072$ & $0.388$ \\ 
    Strength & $-0.327$ & $ 0.219$ & $0.137$ \\ \hline
  \end{tabular}
  \caption{Results of the GLMM analysis with all independent variables for the J-League data when a year was regarded as a season. ${\rm Win}_{t-i}$ ($i=1, 2$) is the binary variable representing whether or not the team has won the ($t-i$)th match ($0$: no win, $1$: win). Likewise for ${\rm Loss}_{t-i}$ ($i=1, 2$) (0: no loss, 1: loss). Home is equal to $0$ for an away game and $1$ for a home game. Strength is equal to the fraction of the matches that the team has won in a season. SE: standard error.}
  \label{table:j_half_glmm2}
  \end{center}
\end{table}

\newpage
\begin{table}[H]
\begin{center}
  \begin{tabular}{cccc}
    Variable & Coefficient & SE & p-value \\ \hline
    Formation change ($\beta_{\rm f}$) & $0.053$ & $0.033$ & $0.109$ \\
    Home ($\beta_{\rm h}$) & $0.134$ & $0.016$ & $< 0.001$ \\
    Win ($\beta_{\rm w}$) & $0.019$ & $0.034$ & $0.572$ \\
    Loss ($\beta_{\rm \ell}$) & $0.079$ & $0.034$ & $0.020$ \\
    Strength ($\beta_{\rm r}$)& $2.645$ & $0.093$ & $< 0.001$ \\ \hline
  \end{tabular}
  \caption{Effects of variables on match results for the J-League data when a year was regarded as a season. The ordered probit model was used. SE: standard error.}
  \label{table:j_half_probit}
  \end{center}
\end{table}

\newpage
\begin{table}[H]
\begin{center}
  \begin{tabular}{cccc}
    Variable & Coefficient & SE & p-value \\ \hline
    ${\rm Win}_{t-1}$ & $-0.482$  & $0.030$ & $< 0.001$ \\
    ${\rm Loss}_{t-1}$ & $0.407$ & $0.031$ & $< 0.001$ \\ \hline
  \end{tabular}
  \caption{Results of the GLMM analysis for the Fussballdaten data when the results of the previous match were used as the sole independent variables. ${\rm Win}_{t-1}$ denotes the binary variable representing whether the team has won the previous match ($0$: no win, $1$: win). Likewise for ${\rm Loss}_{t-1}$ ($0$: no loss, $1$: loss). SE: standard error. }
  \label{table:fussball_glmm1}
  \end{center}
\end{table}

\newpage
\begin{table}[H]
\begin{center}
  \begin{tabular}{cccc}
    Variable & Coefficient & SE & p-value \\ \hline
    ${\rm Win}_{t-1}$ & $-0.568$ & $0.038$ & $< 0.001$ \\
    ${\rm Win}_{t-2}$ & $-0.234$ & $0.038$ & $< 0.001$ \\
    ${\rm Loss}_{t-1}$ & $0.404$ & $0.038$ & $< 0.001$ \\
    ${\rm Loss}_{t-2}$ & $0.018$ & $0.037$ & $0.628$ \\
    ${\rm Win}_{t-1}$ $\times$ ${\rm Win}_{t-2}$ & $0.118$ & $0.056$ & $0.035$ \\
    ${\rm Loss}_{t-1}$ $\times$ ${\rm Loss}_{t-2}$ & $0.071$ & $0.058$ & $0.220$ \\
    Home & $-0.174$ & $0.026$ & $< 0.001$ \\ 
    Strength & $0.026$ & $0.078$ & $0.740$ \\ \hline
  \end{tabular}
  \caption{Results of the GLMM analysis for the Fussballdaten data when all the independent variables were considered. ${\rm Win}_{t-i}$ ($i=1, 2$) is the binary variable representing whether or not the team has won the ($t-i$)th match ($0$: no win, $1$: win). Likewise for ${\rm Loss}_{t-i}$ ($i=1, 2$) (0: no loss, 1: loss). Home is equal to $0$ for an away game and $1$ for a home game. Strength is equal to the fraction of the matches that the team has won in a season. SE: standard error.}
  \label{table:fussball_glmm2}
  \end{center}
\end{table}

\newpage
\begin{table}[H]
\begin{center}
  \begin{tabular}{cccc}
    Variable & Coefficient & SE & p-value \\ \hline
    Formation change ($\beta_{\rm f}$) & $ -0.034$ & $0.014$ & $0.019$ \\
    Home ($\beta_{\rm h}$) & $0.420$ & $0.011$ & $< 0.001$ \\
    Win ($\beta_{\rm w}$) & $ -0.066$ & $0.018$ & $< 0.001$ \\
    Loss ($\beta_{\rm \ell}$) & $ -0.003$ & $0.018$ & $0.858$ \\
    Strength ($\beta_{\rm r}$) & $2.798$ & $0.060$ & $< 0.001$ \\ \hline
  \end{tabular}
  \caption{Effects of variables on match results for the Fussballdaten data. The ordered probit model was used.}
  \label{table:fussball_probit}
  \end{center}
\end{table}

\newpage
\begin{table}[H]
\begin{center}
  \begin{tabular}{ccccc}
    Data set & Variable & Mean & 2.5\% & 97.5\% \\ \hline
    J-League & Formation change ($\beta_{\rm f}$) & $0.004$ & $-0.081$ & $0.088$ \\
    & Home ($\beta_{\rm h}$) & $0.150$ & $0.113$ & $0.186$ \\
    & Win ($\beta_{\rm w}$) & $0.078$ & $0.003$ & $0.152$ \\
    & Loss ($\beta_{\rm \ell}$) & $0.068$ & $-0.007$ & $0.143$ \\ \hline
    Bundesliga & Formation change ($\beta_{\rm f}$) & $-0.029$ & $-0.062$ & $0.003$ \\
    & Home ($\beta_{\rm h}$) & $0.421$ & $0.401$ & $0.442$ \\
    & Win ($\beta_{\rm w}$) & $0.016$ & $-0.020$ & $0.051$ \\
    & Loss ($\beta_{\rm \ell}$) & $0.012$ & $-0.024$ & $0.047$ \\ \hline
  \end{tabular}
  \caption{Effects of variables on match results obtained from the hierarchical Bayesian ordered probit model. }
  \label{table:bayes}
  \end{center}
\end{table}


\newpage
\begin{thebibliography}{10}
\providecommand{\url}[1]{\texttt{#1}}
\providecommand{\urlprefix}{URL }
\expandafter\ifx\csname urlstyle\endcsname\relax
  \providecommand{\doi}[1]{doi:\discretionary{}{}{}#1}\else
  \providecommand{\doi}{doi:\discretionary{}{}{}\begingroup
  \urlstyle{rm}\Url}\fi
\providecommand{\bibAnnoteFile}[1]{%
  \IfFileExists{#1}{\begin{quotation}\noindent\textsc{Key:} #1\\
  \textsc{Annotation:}\ \input{#1}\end{quotation}}{}}
\providecommand{\bibAnnote}[2]{%
  \begin{quotation}\noindent\textsc{Key:} #1\\
  \textsc{Annotation:}\ #2\end{quotation}}
\providecommand{\eprint}[2][]{\url{#2}}

\bibitem{Fudenberg1998}
Fudenberg D, Levine DK (1998) The theory of learning in games.
\newblock MIT press, Cambridge
\input{Fudenberg1998}

\bibitem{Camerer2003}
Camerer C (2003) Behavioral game theory: experiments in strategic interaction.
\newblock Princeton University Press, Princeton
\input{Camerer2003}

\bibitem{Pearce2013}
Pearce JM (2013) Animal learning and cognition: an introduction.
\newblock Psychology Press, Hove
\input{Pearce2013}

\bibitem{Sutton1998}
Sutton RS, Barto AG (1998) Reinforcement learning: an introduction.
\newblock MIT press, Cambridge
\input{Sutton1998}

\bibitem{Schultz1997}
Schultz W, Dayan P, Montague PR (1997) A neural substrate of prediction and
  reward.
\newblock Science 275(5306):1593--1599.
\input{Schultz1997}

\bibitem{Glimcher2009}
Glimcher PW, Camerer C, Fehr E, Poldrack RA (2009) Neuroeconomics: decision
  making and the brain.
\newblock Academic Press, New York
\input{Glimcher2009}

\bibitem{Kraines1989}
Kraines D, Kraines V (1989) Pavlov and the prisoner's dilemma.
\newblock Theory Decis 26(1):47--79.
\input{Kraines1989}

\bibitem{Nowak1993}
Nowak M, Sigmund K (1993) A strategy of win-stay, lose-shift that outperforms
  tit-for-tat in the Prisoner's Dilemma game.
\newblock Nature 364(6432):56--58.
\input{Nowak1993}

\bibitem{Wedekind1996}
Wedekind C, Milinski M (1996) Human cooperation in the simultaneous and the
  alternating Prisoner's Dilemma: Pavlov versus Generous Tit-for-Tat.
\newblock Proc Natl Acad Sci USA 93(7):2686--2689.
\input{Wedekind1996}

\bibitem{Milinski1998}
Milinski M, Wedekind C (1998) Working memory constraints human cooperation in
  the Prisoner's Dilemma.
\newblock Proc Natl Acad Sci 95(23):13755--13758.
\input{Milinski1998}

\bibitem{Hayden2009}
Hayden BY, Platt ML (2009) Gambling for Gatorade: risk-sensitive decision
  making for fluid rewards in humans.
\newblock Anim Cogn 12(1):201--207.
\input{Hayden2009}

\bibitem{Scheibehenne2011}
Scheibehenne B, Wilke A, Todd PM (2011) Expectations of clumpy resources
  influence predictions of sequential events.
\newblock Evol Human Behav 32(5):326--333.
\input{Scheibehenne2011}

\bibitem{Mesoudi2008a}
Mesoudi A, O'Brien MJ (2008) The cultural transmission of Great Basin
  projectile-point technology {I}: an experimental simulation.
\newblock Am Antiq 73(1):3--28.
\input{Mesoudi2008a}

\bibitem{Mesoudi2008b}
Mesoudi A, O'Brien MJ (2008) The cultural transmission of Great Basin
  projectile-point technology {II}: an agent-based computer simulation.
\newblock Am Antiq 73(4):627--644.
\input{Mesoudi2008b}

\bibitem{Mesoudi2014}
Mesoudi A (2014) Experimental studies of modern human social and individual
  learning in an archaeological context: people behave adaptively, but within
  limits.
\newblock In: Akazawa T, Ogihara N, C~Tanabe HC, Terashima H (ed) Dynamics
  of learning in Neanderthals and modern humans Vol 2, Springer, Heidelberg, pp 65--76.
\input{Mesoudi2014}

\bibitem{Vyse2013}
Vyse SA (2013) Believing in magic: the psychology of superstition updated
  edition.
\newblock Oxford University Press, Oxford
\input{Vyse2013}

\bibitem{Panja2013}
Panja T (2013)
\newblock {Top soccer leagues get 25\% rise in TV rights sales, report says}
  {In: Bloomberg Business.} 
\newblock
  \url{http://www.bloomberg.com/news/articles/2013-11-11/top-soccer-leagues-get-25-rise-in-tv-rights-sales-report-says}.
\newblock Accessed 28 February 2015.
\input{Panja2013}

\bibitem{Dobson2011}
Dobson S, Goddard J (2001) The economics of football.
\newblock Cambridge University Press, Cambridge
\input{Dobson2011}

\bibitem{Dixon1997}
Dixon MJ, Coles SG (1997) Modelling association football scores and
  inefficiencies in the football betting market.
\newblock J R Stat Soc, Ser C, Appl Stat 46(2):265--280.
\input{Dixon1997}

\bibitem{Wilson2013}
Wilson J (2013) Inverting the pyramid: the history of soccer tactics.
\newblock Nation Books, New York
\input{Wilson2013}

\bibitem{Audas1997}
Audas R, Dobson S, Goddard J (1997) Team performance and managerial change in
  the english football league.
\newblock Econ Aff 17(6):30--36.
\input{Audas1997}

\bibitem{Dawson2000}
Dawson P, Dobson S, Gerrard B (2000) Estimating coaching efficiency in
  professional team sports: evidence from english association football.
\newblock Scot J Polit Econ 47(4):399--421.
\input{Dawson2000}

\bibitem{Audas2002}
Audas R, Dobson S, Goddard J (2002) The impact of managerial change on team
  performance in professional sports.
\newblock J Econ Bus 54(6):633--650.
\input{Audas2002}

\bibitem{Bangsbo2000}
Bangsbo J, Peitersen B (2000) {Soccer systems \& strategies}.
\newblock Human Kinetics, Champaign
\input{Bangsbo2000}

\bibitem{Hirotsu2006}
Hirotsu N, Wright MB (2006) Modeling tactical changes of formation in
  association football as a zero-sum game.
\newblock J Quant Anal Sports 2(2):4.
\input{Hirotsu2006}

\bibitem{JLeague}
{J-League Data site}.
\newblock \url{https://data.j-league.or.jp/SFTP01/}.
\newblock Accessed 4-6 June 2014.
\input{JLeague}

\bibitem{Kicker}
{Kicker-online}.
\newblock \url{http://www.kicker.de/}.
\newblock Accessed 14-16 Aug 2014.
\input{Kicker}

\bibitem{Fussballdaten}
{Fussballdaten}.
\newblock \url{http://www.fussballdaten.de/}.
\newblock Accessed 28-31 July 2014.
\input{Fussballdaten}

\bibitem{Goh2008}
Goh KI, Barab\'{a}si AL (2008) Burstiness and memory in complex systems.
\newblock EPL 81:48002.
\input{Goh2008}

\bibitem{R2014}
{R Core Team} (2014) R: A Language and Environment for Statistical Computing.
\newblock R Foundation for Statistical Computing, Vienna.
\newblock \urlprefix\url{http://www.R-project.org/}.
\input{R2014}

\bibitem{lme4}
Bates D, Maechler M, Bolker B, Walker S (2014) {lme4}: Linear mixed-effects
  models using Eigen and S4.
\newblock \urlprefix\url{http://CRAN.R-project.org/package=lme4}.
\newblock R package version 1.1-7.
\input{lme4}

\bibitem{Koning2000}
Koning RH (2000) {Balance in competition in Dutch soccer}.
\newblock J R Stat Soc, Ser D, The Statistician 49:419--431.
\input{Koning2000}

\bibitem{Parent2012}
Parent E, Rivot E (2012) Introduction to hierarchical Bayesian modeling for
  ecological data.
\newblock CRC Press, Florida
\input{Parent2012}

\bibitem{maxLik}
Henningsen A, Toomet O (2011) {maxLik}: A package for maximum likelihood
  estimation in {R}.
\newblock Comp Stat 26(3):443-458.
\input{maxLik}

\bibitem{Albert2010}
Albert J, Koning RH (2010) {Statistical thinking in sports}.
\newblock CRC Press, Florida
\input{Albert2010}

\bibitem{Taiji1999}
Taiji M, Ikegami T (1999) Dynamics of internal models in game players.
\newblock Physica D 134(2):253--266.
\input{Taiji1999}

\bibitem{Macy2002}
Macy MW, Flache A (2002) Learning dynamics in social dilemmas.
\newblock Proc Natl Acad Sci USA 99(Suppl 3): 7229--7236.
\input{Macy2002}

\bibitem{Masuda2009}
Masuda N, Ohtsuki H (2009) A theoretical analysis of temporal difference
  learning in the iterated prisoner's dilemma game.
\newblock Bull Math Biol 71(8):1818--1850.
\input{Masuda2009}

\bibitem{Masuda2011}
Masuda N, Nakamura M (2011) Numerical analysis of a reinforcement learning
  model with the dynamic aspiration level in the iterated prisoner's dilemma.
\newblock J Theor Biol 278(1):55--62.
\input{Masuda2011}

\bibitem{Neiman2011}
Neiman T, Loewenstein Y (2011) Reinforcement learning in professional
  basketball players.
\newblock Nat Commun 2:569.
\input{Neiman2011}

\bibitem{Bar-Eli2006}
Bar-Eli M, Avugos S, Raab M (2006) {Twenty years of ``hot hand'' research:
  review and critique}.
\newblock Psychol Sport Exerc 7(6):525--553.
\input{Bar-Eli2006}

\bibitem{Bialkowski2014a}
Bialkowski A, Lucey P, Carr P, Yue Y, Matthews I (2014) {``Win at home and draw
  away'': automatic formation analysis highlighting the differences in home and
  away team behaviors}.
\newblock In: MIT Sloan Sports Analytics Conference (SSAC).
\input{Bialkowski2014a}

\bibitem{Bialkowski2014b}
Bialkowski A, Lucey P, Carr P, Yue Y, Sridharanand S, et~al. (2014) Large-scale
  analysis of soccer matches using spatiotemporal data.
\newblock In: IEEE International Conference on Data Mining (ICDM).
\input{Bialkowski2014b}

\bibitem{RStan2014}
{Stan Development Team} (2014).
\newblock Rstan: the {R} interface to stan, version 2.5.0.
\newblock \urlprefix\url{http://mc-stan.org/rstan.html}.
\input{RStan2014}


\end{thebibliography}
\end{document}